\documentstyle[12pt]{article}
\begin{document} 
\newcommand{\be}{\begin{equation}} \newcommand{\ee}{\end{equation}}
\begin{titlepage} 
\title{Light amplification by gravitational waves in scalar--tensor theories 
of gravity} 
\author{Valerio Faraoni \\ \\{\small \it
Department of Physics and Astronomy, University of Victoria} \\ 
{\small \it
P.O. Box 3055, Victoria, B.C. V8W 3P6 (Canada)}} 
\date{}
\maketitle   \thispagestyle{empty}  \vspace*{1truecm}
\begin{abstract} 
It is shown that the amplification of a light beam by
gravitational waves in scalar--tensor theories of gravity is a first order
effect in the wave amplitudes. In general relativity, instead, the effect 
is only of second order. 
\end{abstract} 
\vspace*{1truecm} 
\begin{center} {\small 
To appear in {\em Astrophysical Letters and Communications}
}  \end{center}  \end{titlepage}            \clearpage 

Scalar--tensor theories (ST) of gravity have been considered as 
alternatives to
general relativity (GR) for a long time (see e.g. Damour and Esposito--Farese
1992, and references therein). Recent interest in ST theories arises from the 
facts that in 
supergravity and superstring theories, scalar 
fields are associated to the usual tensor field describing
gravity (Green {\em et al.} 1987), and that a coupling between a scalar 
field and gravity seems unavoidable in string theories (Casas {\em et al.}
1991). Further motivation for the study of ST theories comes from the 
extended and hyperextended inflationary scenarios (La and Steinhardt 1989;
Accetta and Steinhardt 1990; Steinhardt 1993). 

In this Letter, we consider the amplification of a light beam by 
gravitational waves in ST theories. This phenomenon was studied in the context 
of GR, 
but it was discarded as an academic topic because its magnitude 
is of second order in the 
waves' amplitude, and therefore exceedingly small 
(Zipoy 1966; Zipoy and Bertotti 1968; Bertotti 1971; Bertotti and 
Catenacci 1975). New interest in lensing by gravitational waves in 
GR arose recently (Faraoni 1992{\em a,~b}; Labeyrie 1993; Durrer 1994; 
Fakir 1993, 1994{\em a,~b,~c}, 1995). 
In the following, we show that in ST theories the amplification effect is 
of {\em first order} in the wave amplitudes, and hence may 
be of interest in astrophysics. The difference with respect to GR 
is due to the fact that in ST theories gravitational waves have a spin~0 
polarization mode, in addition to the usual spin~2 modes (which are the 
only present in GR).

We consider the ST theories of gravity described by the action \footnote{The
metric signature is $+2$. Greek indices run from 0 to 3 and Latin indices run
from 1 to 3. A comma and a semicolon denote, respectively, ordinary and
covariant differentiation. The Ricci tensor is given by
$R_{\mu\nu}=
\Gamma^{\sigma}_{\mu\nu ,\sigma} -\Gamma^{\sigma}_{\mu\sigma ,\nu}+
\Gamma^{\rho}_{\mu\nu}\Gamma^{\sigma}_{\rho\sigma}-
\Gamma^{\sigma}_{\rho\nu}\Gamma^{\rho}_{\mu\sigma} $. Units in 
which the Newton constant and the speed of light assume the 
value unity are adopted. Round [square] brackets denote 
[anti]symmetrization.}
\be    \label{1} 
S=\frac{1}{16 \pi} \int d^4x\sqrt{-g}\left[ R\phi
-\frac{\omega ( \phi)}{\phi}\, \nabla^{\alpha}\phi \nabla_{\alpha} \phi 
\right] +S_{ng} \; , 
\ee 
where $R$ is the Ricci curvature, $\phi$ is the cosmological scalar field, 
$g$ is the determinant of the metric and $S_{ng}$ is the nongravitational 
part of the action, which we will ignore in what follows. This is 
not the most general ST theory, since it is possible to replace the 
term $R \phi $ in the Lagrangian density of Eq.~(\ref{1}) with 
$ R f( \phi) $, where $f$ is an arbitrary function 
of $\phi$, and to include a cosmological term $\Lambda ( 
\phi)$. However, the action~(\ref{1}) is sufficiently 
general for our purposes. We refer the reader to the book by Will (1993) for 
the field equations of the theory, and we consider local perturbations 
around a flat 
background. We assume that the spacetime metric and the scalar
field are given by 
\be   \label{5} 
g_{\mu\nu}=\eta_{\mu\nu}+h_{\mu\nu} \; , 
\ee
\be \label{6} 
\phi=\phi_0+\varphi  \; , 
\ee 
in a Cartesian coordinate system, where $\eta_{\mu\nu}=$diag($-1, 1, 1, 1$), 
$\phi_0$ is constant, and the order of magnitude of the perturbations is 
$ \left| h_{\mu\nu} \right| \sim \left| \varphi /\phi_0  
\right|=\mbox{O}( \epsilon ) $ in terms of a smallness parameter 
$\epsilon \ll 1$.
Introducing the quantity $ \Theta_{\mu\nu}\equiv
h_{\mu\nu}-\frac{1}{2} \,\eta_{\mu\nu} h-\eta_{\mu\nu} 
\varphi/\phi_0 $
and imposing the transversality condition 
$ \partial^{\nu} \Theta_{\mu\nu}=0 $,
the linearized field equations in vacuum for the perturbations are 
derived (Will 1993):
\begin{eqnarray}     
&& \Box \Theta_{\mu\nu}=0 \; , \label{9} \\
&& \Box \varphi=0  \; . \label{10} 
\end{eqnarray}
To first order, one has (Will 1993) 
\be   \label{13}
R_{\mu\nu}=\frac{\partial_{\mu}\partial_{\nu}\varphi}{\phi_0} 
\ee 
and $ R=0 $. The solutions of Eq.~(\ref{10}) are expressed as Fourier 
integrals of plane waves 
\be  
\label{14} \varphi=\varphi_0 \exp{\left( \pm \, i \,l_{\alpha} 
x^{\alpha}\right)} \; , 
\ee
where $\varphi_0$ is a constant and $l^{\mu}$ is a constant vector 
satisfying $\eta_{\mu\nu}l^{\mu}l^{\nu}=0$. 
In ST theories, a plane monochromatic wave propagating along a definite 
direction has a 
spin~0 (``T0'') polarization mode associated to the scalar 
field $\varphi$, in
addition to the usual spin~2 modes associated to the tensor field
$h_{\mu\nu}$ and familiar from GR 
(Robinson and Winicour 1969; Wagoner 1970). Therefore, ST 
theories described by 
the action~(\ref{1}) are of class $N_3$ in the {\em E(2)} classification 
scheme of Eardley {\em et al.} (1973{\em a,~b}). 
The presence of the spin~0 mode in ST 
gravitational waves is responsible for the differences in the 
amplification effect in ST theories and in GR.

The amplification of a light beam crossing gravitational waves was studied 
in various papers in the 
context of GR (Sachs 1961; Zipoy 1966; Bertotti 1971). Following Bertotti
(1971), we consider a congruence of null
rays propagating from a light source to an observer in the geometric optics
approximation. Let $k^{\mu}$ be the tangent field to the null geodesics: the
optical scalars and the complex shear of the congruence are defined 
by 
\begin{eqnarray}    
&& \theta=\frac{1}{2} \,
{k^{\alpha}}_{;\alpha} \; , \label{15} \\
&&  \left| \sigma \right|^2=\frac{1}{2} \, k_{( \alpha;\beta)} 
k^{\alpha;\beta}-\theta^2 \; , \label{16} \\
&& \omega^2=\frac{1}{2} \, k_{[\alpha;\beta]} \, k^{\alpha;\beta} 
\: ,\label{17}
\end{eqnarray} 
\be   \label{18}
\sigma=k_{\alpha;\beta} \, 
\bar{t}^{\alpha}t^{\beta}=\frac{1}{\sqrt{2}} \left[
k_{( \alpha;\beta)}k^{\alpha;\beta}-\frac{1}{2} \left( 
{k^{\alpha}}_{;\alpha}
\right)^2 \right]^{1/2} \; ,
\ee 
respectively (Sachs 1961). $t^{\mu}$ is a complex vector satisfying 
$ t^{\mu}t_{\mu}=k^{\mu}t_{\mu}=0$, $t_{\alpha}\bar{t}^{\alpha}=-1$ (a bar 
denoting complex conjugation). The
optical scalars satisfy the propagation equations 
\begin{eqnarray}   
&& \frac{d\theta}{d\lambda}=-\,\theta^2-|\sigma|^2+\omega^2- \frac{1}{2} \,
R_{\mu\nu}k^{\mu}k^{\nu} \; , \label{19} \\
&& \frac{d\sigma}{d\lambda}=-2
\, \theta \, \sigma-C_{\mu\nu\rho\tau}\bar{t}^{\mu}
k^{\nu}\bar{t}^{\rho}k^{\tau}\equiv -2\, \theta \, \sigma -C( \lambda) \; , 
\label{20} \\
&& \frac{d\omega}{d\lambda}=-2 \, \omega \, \theta \; , \label{21} 
\end{eqnarray}
(Sachs 1961), where $ \lambda$ is an affine parameter along the 
null geodesics and 
${C^{\mu}}_{\nu\rho\tau}$ is the Weyl tensor. For ease of comparison with the
discussion of Bertotti (1971) in the context of GR, we restrict to 
the case in which the spacetime
metric and the scalar field are given\footnote{The
physically more relevant situation in which light propagates through a
realistic cosmological model exhibits some complications that are unnecessary
in order to understand the physics of the effect and can be introduced at a
later stage. In this respect, our approach is 
similar to that of conventional gravitational lens theory 
(Schneider {\em et al.} 1992).} by Eqs.~(\ref{5}) and (\ref{6}). 

In GR, with the metric~(\ref{5}) and gravitational
waves of amplitudes $h_{\mu\nu}=$O($\epsilon$), the Ricci tensor vanishes.
The solution to Eqs.~(\ref{19})--(\ref{21}), to the lowest order in 
$\epsilon$ and for a
point--like source of light, is then given by 
$ \theta=\lambda^{-1}+\delta \theta $, $\sigma=\delta\sigma$, $\omega=0 $ 
(Sachs 1961), in the approximation of small $\delta\theta$ and $\delta
\sigma$ (para--axial approximation) and with the convention 
that $\lambda=0$ at the position of the
light source. $\delta \theta$ and $\delta\sigma$ obey the equations 
\begin{eqnarray}
&& \frac{1}{\lambda^2} \frac{d}{d\lambda} \left( \lambda^2 \delta\theta
\right)=-\,\delta\theta^2-\left|\delta\sigma \right|^2   \; ,   \label{25} \\
&& \frac{1}{\lambda^2} \frac{d}{d\lambda} \left( \lambda^2 \delta\sigma
\right)=-2 \, \delta \theta \, \delta\sigma -C( \lambda ) \; . \label{26} 
\end{eqnarray}
From Eqs.~(\ref{25}) and (\ref{26}) it was deduced that the perturbations in 
the shear and the expansion of the 
congruence of null rays have orders of magnitude $\delta 
\sigma=$O$( \epsilon)$ and $\delta
\theta=$O($ \epsilon^2$) (Bertotti 1971). Thus, the amplification of a pencil
of light rays by gravitational waves in GR vanishes to 
first order, reducing to
a second order effect. The same result was obtained by Zipoy (1966) with a 
different method.

The optical scalars formalism is independent of the field equations of the
theory and can be applied also in ST theories. For gravitational waves 
in ST theories, the Ricci 
tensor is not identically zero, but is given by Eq.~(\ref{13}). For the simple
monochromatic plane wave~(\ref{14}), Eq.~(\ref{25}) is replaced by 
\be   \label{27} 
\frac{1}{\lambda^2}\frac{d}{d\lambda}\left( \lambda^2
\delta\theta \right)=-\delta\theta^2-|\delta\sigma|^2+ \frac{1}{2}\left(
l_{\mu}k^{\mu}\right)^2 \frac{\varphi}{\phi_0} \; ,
\ee 
where the right hand side is dominated by its last term. Equation (\ref{27}) 
implies that the expansion of the congruence of null rays is 
$\delta\theta=$O$(\varphi/\phi_0 )=$O$( \epsilon)$ when
$l_{\mu}k^{\mu}\neq 0$. Thus, the amplification of a pencil of light rays is a
{\em first order} effect in ST theories, contrarily to GR. The effect 
disappears (to this order) when $l^{\mu}$ and $k^{\mu}$ are 
parallel, due to the
transversality of ST gravitational waves (expressed by the equation 
$ \partial^{\nu} \Theta_{\mu\nu}=0 $). The sign
of the Ricci term in Eq.~(\ref{27}) oscillates in time with the period of 
the ST wave, corresponding to an alternation of focusing and defocusing 
of the beam (amplification and attenuation of the image of the light source).
Equation~(\ref{27}) shows clearly that the difference between GR and ST 
theories consists in the presence of matter (i.e. the scalar field) in the 
path of the light
beam in the latter case\footnote{If one describes the ST theory in the 
``Einstein frame'' conformally
related to the ``Jordan frame'' that we are using (see Magnano and Sokolowski
(1994) and references therein for the definition of this terminology and 
a discussion), the Lagrangian in the new frame contains a new scalar 
field which behaves again as matter in the light
beam, giving a first order amplification effect also in the Einstein frame.}, 
while in the GR case lensing is due only to 
the shear of the gravitational field (compare the description of an ordinary
gravitational lens given in Schneider {\em et al.} (1992) using the 
vector formalism).

The previous argument suggests that the amplification effect in ST theories
could be relevant for astrophysics and perhaps 
detectable by observations, thus
providing a means to distinguish between a ST theory and GR. 
Unfortunately, whether this is or not the case cannot be decided on the basis
of the approach used here. In fact, the previous analysis is based on the 
optical scalars formalism,
which is known to fail even in the qualitative description of many 
interesting aspects of gravitational lensing, and is not suitable for a
quantitative analysis (Schneider {\em et al.} 1992). 
A formalism
capable of describing lensing by gravitational waves is not available at
present, although progress is being made in this direction 
(Faraoni 1992{\em a,~b}; work in progress). This case is more complicated
than lensing by ordinary mass distributions because of the difficulties
encountered in applying Fermat's principle to a non--(conformally) stationary
spacetime. The lack of literature on this subject is probably due to the fact
that the amplification of light was believed 
to be negligible for gravitational
waves in GR. It is hoped that the observation that the effect is much 
stronger in 
ST theories than in GR will stimulate further research on 
this subject. 

{\small \section*{Acknowledgments} 

This work was supported by the NATO Advanced Fellowships Programme through the
National Research Council of Italy (CNR). The author acknowledges the warm 
hospitality at the University of Victoria. 

\section*{References} 

\noindent Accetta, F. S. and Steinhardt, P. J. 1990, {\em Phys. 
Rev. Lett.} {\bf 64}, 2740.\\
\noindent Bertotti, B. 1971, in {\em General Relativity and
Cosmology, XLII Course of the Varenna Summer School}, R. K. Sachs ed., 
Academic Press, New York, p.~347.\\
\noindent Bertotti, B. and Catenacci, R. 1975, {\em Gen. 
Rel. Grav.} {\bf 6}, 329.\\
\noindent Casas, J. A., Garcia--Bellido, J. and Quir\'{o}s, M. 1991, 
{\em Nucl. Phys. B} {\bf 361}, 713.\\
\noindent Damour, T. and Esposito--Farese, G., 1992, {\em Class. Quant. Grav.} 
{\bf 9}, 2093.\\
\noindent Durrer, R. 1994, {\em Phys. Rev. Lett.} {\bf 72}, 3301. \\
\noindent Eardley, D. M., Lee, D. L., Lightman, A. P., Wagoner, R. V. and 
Will, C. M. 1973{\em a}, {\em Phys. Rev. Lett.} {\bf 30}, 884.\\
\noindent Eardley, D. M., Lee, D. L. and Lightman, A. P. 1973{\em b}, {\em 
Phys. Rev. D} {\bf 8}, 3308.\\
\noindent Fakir, R. 1993, {\em Ap. J.} {\bf 418}, 202.\\
\noindent ------------ 1994{\em a}, {\em Phys. Rev. D} {\bf 50}, 3795. \\
\noindent ------------ 1994{\em b}, {\em Ap. J.} {\bf 426}, 74. \\
\noindent ------------ 1994{\em c}, preprint UBCTP--94--011. \\
\noindent ------------ 1995,  preprint astro--ph~9507112. \\
\noindent Faraoni, V. 1992{\em a}, in {\em Gravitational Lenses}, Proceedings,
Hamburg 1991, R. Kayser, T. Schramm and L. Nieser eds., Springer--Verlag, 
Berlin.\\
\noindent Faraoni, V. 1992{\em b}, {\em Ap. J.} {\bf 398}, 425. \\
\noindent Green, B., Schwarz, J. M.  and Witten, E. 1987, {\em
Superstring Theory}, Cambridge University Press, Cambridge.\\
\noindent La, D. and Steinhardt, P. J. 1989, {\em Phys. Rev. 
Lett.} {\bf 62}, 376. \\
\noindent Labeyrie, A. 1993, {\em Astron. Astrophys.} {\bf 268}, 823.\\
\noindent Magnano, G. and Sokolowski, L. M. 1994, {\em Phys. Rev. D} {\bf 50}, 
5039. \\
\noindent Robinson D. C. and Winicour, J. 1969, {\em Phys. Rev. Lett.} 
{\bf 22}, 198.\\
\noindent Sachs, R. K. 1961, {\em Proc. Roy. Soc. Lond. A} {\bf 264}, 309.\\
\noindent Schneider, P., Ehlers, J. and Falco, E. E. 1992, {\em Gravitational 
Lenses}, Springer--Verlag, Berlin.\\
\noindent Steinhardt, P. J. 1993, {\em Class. Quant. Grav.} 
{\bf 10}, S33.\\
\noindent Zipoy, D. M. 1966, {\em Phys. Rev.} {\bf 142}, 825.\\
\noindent Zipoy, D. M. and Bertotti, B. 1968, {\em Nuovo 
Cimento} {\bf 56B}, 195.\\
\noindent Wagoner, R. V. 1970, {\em Phys. Rev. D} {\bf 1}, 3209.\\
\noindent Will, C. M. 1993, {\em Theory and Experiment in Gravitational
Physics} (revised edition), Cambridge University Press, Cambridge.  }
\end{document}